\documentclass[12pt]{article}
\usepackage{graphicx, amssymb, amsmath, multicol, url, epsfig}
\usepackage[english]{babel}
\catcode`\«=\active
\catcode`\»=\active
\def«{\og\ignorespaces}%
\def»{{\fg}}%

\title{ A result on the phase diagram of a Ginzburg-Landau problem}
\def\QuotS#1#2{\leavevmode\kern-.0em\raise.2ex\hbox{$#1$}\kern-.1em/\kern-.1em\lower.25ex\hbox{$#2$}}
\DeclareMathAlphabet{\mathsfsl}{OT1}{cmss}{m}{n}
\DeclareMathAlphabet{\mathbfsl}{OT1}{pss}{b}{n}

\author{Mathieu Dutour\\
ENS, Paris and Hebrew University, Jerusalem,\footnote{Research financed by EC's IHRP Programme, within the Research Training Network ``Algebraic Combinatorics in Europe,'' grant HPRN-CT-2001-00272.}
}

\date{}
\begin{document}
\def\Tore#1{\leavevmode\kern-.0em\raise.2ex\hbox{$\R^2$}\kern-.1em/\kern-.1em\lower.25ex\hbox{$#1$}}

\newcommand{\D}{\displaystyle}
\newcommand{\R}{\ensuremath{\mathbb{R}}}
\newcommand{\N}{\ensuremath{\mathbb{N}}}
\newcommand{\Q}{\ensuremath{\mathbb{Q}}}
\newcommand{\C}{\ensuremath{\mathbb{C}}}
\newcommand{\Z}{\ensuremath{\mathbb{Z}}}
\newcommand{\T}{\ensuremath{\mathbb{T}}}
\newcommand{\K}{\ensuremath{\mathbb{K}}}
\newcommand{\curl}{\ensuremath{\mathop{\rm curl}\nolimits}}
\newcommand{\divergence}{\ensuremath{\mathop{\rm div}\nolimits}}
\newcommand{\determinant}{\ensuremath{\mathop{\rm det}\nolimits}}
\newcommand{\dx}{\ensuremath{\mathop{\rm dx}\nolimits}}
\newcommand{\Rez}{\ensuremath{\mathop{\rm Re}\nolimits}}
\newcommand{\Imz}{\ensuremath{\mathop{\rm Im}\nolimits}}

\newtheorem{theorem}{Theorem}
\newtheorem{lemma}[theorem]{Lemma}
\newtheorem{proposition}[theorem]{Proposition}
\newtheorem{definition}[theorem]{Definition}
\newtheorem{remark}[theorem]{Remark}
\newtheorem{corollary}[theorem]{Corollary}
\newtheorem{exemple}[theorem]{Exemple}
\newtheorem{problem}[theorem]{Problem}
\newtheorem{hypothesis}[theorem]{Hypothesis}
\newcommand{\qed}{\hfill $\Box$ }
\newcommand{\proof}{\noindent{\bf Proof.}\ \ }

\maketitle
\begin{abstract}
\noindent Working with a particular modelization of Ginzburg-Landau phenomenological theory (see \cite{dutourII}, \cite{dutour} and Section \ref{change-var}), we first recall the form of the phase diagram of this modelization as it usually drawn in the physical literature (\cite{tink}, \cite{Kittel}, \cite{sarma} and \cite{PG-de-Gennes}).

We then study in detail the special case, when the critical Ginzburg Landau parameter $k$ is equal to $\frac{1}{\sqrt{2}}$.
This allows us to prove that the critical magnetic field $H_{c1}(k)$ is strictly decreasing at $k=\frac{1}{\sqrt{2}}$.

\end{abstract}

\section[I]{INTRODUCTION}\label{Introduction}
\noindent In 1950 V. Ginzburg and L. Landau (\cite{GL}) have proposed a modelization for describing the various states of a superconducting material. They introduce a functional depending on a {\em wave function} $\phi$ and a {\em magnetic potential vector} $\mathbf{A}$, whose local minima will describe the properties of the material; in this modelization $|\phi|^2$ represents the local density of superconducting electrons.

Abrikosov (\cite{Abrikosov}) has introduced a particular Ginzburg-Landau modelization, which predicts the periodic structure for the zeros of $\phi$, which was subsequently observed in experiments.
His model depends on two positive parameters $k$ and $H_{\mathsfsl{ext}}$, called {\em Ginzburg-Landau parameter} and {\em external magnetic field}. It also assumed that:
\begin{enumerate}
\item The superconductor is infinite, homogeneous and isotrop.
\item The magnetic field $\mathbf{H_{\mathsfsl{ext}}}=(0, 0, H_{\mathsfsl{ext}})$ is constant.
\item The energy functional $F(\phi, \mathbf{A})$ has a Ginzburg-Landau form and depends on the {\em Ginzburg-Landau parameter} $k$.
\item The pairs $(\phi, \mathbf{A})$ considered are gauge invariant along the z-axis and also along a lattice of $\R^2$.
\item The lattice has a fixed shape and there is one quantum flux per unit cell of it.
\end{enumerate}
After some change of variable, recalled in Section \ref{change-var}, we obtain the following formulation of the problem:

Denote ${\cal L}$ a lattice of $\R^2$, with fundamental domain $\Omega$ of area $1$. Define the vector bundle $E_1$ over $\QuotS{\R^2}{\cal L}$ as the vector bundle, whose $C^{\infty}$ sections are described by
\begin{equation*}\label{definition-du-fibre}
C^{\infty}(E_1)=\left\{
\begin{array}{c}
u:\R^2\rightarrow\C\mbox{~s.t.~}\forall (x,y)\in\R^2, \forall v=(v_x,v_y)\in{\cal L},\\
u((x,y)+v)=e^{i\pi(v_{x}y-v_{y}x)}u(x,y)
\end{array}\right\}
\end{equation*}
The vector bundle $E_1$ is non-trivial; this implies that any section $u\in C^{\infty}(E_1)$ has at least one zero in $\QuotS{\R^2}{\cal L}$.

The potential vector $\mathbf{a}$ belongs to the space
\begin{equation*}\label{VectorBundle-for-a}
\{\mathbf{a}\in H^1_{\mathsfsl{loc}}(\R^2; \R^2)\mbox{~such~that~}\divergence\,\mathbf{a}=0,\,\,\mathbf{a}\mbox{~is~${\cal L}$-periodic~and~}\int_{\Omega}\mathbf{a}=0\}
\end{equation*}
We denote by ${\cal A}$ the space of all pairs $(u,\mathbf{a})$ with $u$ being a $H^1_{\mathsfsl{loc}}$ section of $E_1$ and $\mathbf{a}$ belonging to the above space.

Denote $H_{\mathsfsl{int}}$ the {\em internal magnetic field} and $E_{k,H_{\mathsfsl{int}}}$ the functional defined over ${\cal A}$ by
\begin{equation*}\label{definition-Emuk}
\begin{array}{rcl}
E_{k,H_{\mathsfsl{int}}}(u,\mathbf{a})
&=&\int_{\Omega}\frac{\mu}{2}\Vert i\mathbf{\nabla}u+(\mathbf{A}_{0}+\mathbf{a})u\Vert^2+\frac{1}{4}(1-|u|^2)^2+\frac{\mu^2k^2}{2}|\curl\, \mathbf{a}|^2
\end{array}
\end{equation*}
with $\mu=\frac{H_{\mathsfsl{int}}}{2\pi k}$ and $\mathbf{A}_0=\pi{\scriptstyle \left(\begin{array}{c}
-y\\
x
\end{array}\right)}$. We then define the energy of the superconductor as
\begin{equation*}
E_{k, H_{\mathsfsl{ext}}}(H_{\mathsfsl{int}}, u, \mathbf{a})=E_{k,H_{\mathsfsl{int}}}(u,\mathbf{a})+\frac{1}{2}(H_{\mathsfsl{int}}-H_{\mathsfsl{ext}})^2 .
\end{equation*}
The term $\frac{1}{2}(H_{\mathsfsl{int}}-H_{\mathsfsl{ext}})^2$ is a simple magnetic energy, while the term $E_{k,H_{\mathsfsl{int}}}$ is the internal energy of the superconductor.
The energy ${\cal E}_{k,H_{\mathsfsl{ext}}}$ is then defined as the minimum of $E_{k,H_{\mathsfsl{ext}}}$ over all magnetic field $H_{\mathsfsl{int}}$ and pairs $(u,\mathbf{a})\in {\cal A}$. Also we denote $m_E(k,H_{\mathsfsl{int}})$ the infimum of $E_{k,H_{\mathsfsl{int}}}$ over all pairs $(u,\mathbf{a})\in {\cal A}$.

For $u=0$, $\mathbf{a}=0$ and $H_{\mathsfsl{int}}=H_{\mathsfsl{ext}}$ one obtains the energy $E_{\cal N}=\frac{1}{4}$, which is the energy of the so called {\em normal state}. In the limit case $H_{\mathsfsl{int}}=0$, one obtains (see \cite{dutour} or \cite{dutourII}) the energy $E_{\cal P}=\frac{H_{\mathsfsl{ext}}^2}{2}$, which is the energy of the {\em pure state}.
This leads us to introduce three sets in $\R_+^*\times \R_+^*$:
\begin{equation*}\label{Definition-Trois-ensembles}
\begin{array}{rcl}
{\cal N}&=&\{(k, H_{\mathsfsl{ext}})\in \R_+^*\times\R_+^*\mbox{~s.t.~} {\cal E}_{k, H_{\mathsfsl{ext}}}=E_{\cal N}\}\;,\\
{\cal P}&=&\{(k, H_{\mathsfsl{ext}})\in \R_+^*\times\R_+^*\mbox{~s.t.~} {\cal E}_{k, H_{\mathsfsl{ext}}}=E_{\cal P}\}\;,\\
{\cal M}&=&\{(k, H_{\mathsfsl{ext}})\in \R_+^*\times\R_+^*\mbox{~s.t.~} {\cal E}_{k, H_{\mathsfsl{ext}}}<\inf(E_{\cal P}, E_{\cal N})\}\;.
\end{array}
\end{equation*}
The set ${\cal M}$ is the complementary of ${\cal P}\cup {\cal N}$ in $\R_+^*\times \R_+^*$; if $(k,H_{\mathsfsl{ext}})\in {\cal M}$, then the superconductor is said to be in a {\em mixed state}.

Using this simple modelization we were able (see \cite{dutour} and \cite{dutourII}) to prove following {\em monotonicity theorem}.

\begin{theorem}\label{TheoremMonotony}
(i) If $(k, H_{\mathsfsl{ext}})\in {\cal P}$, $k'\leq k$ and $H'_{ext}\leq H_{\mathsfsl{ext}}$ then $(k',H'_{ext})\in {\cal P}$.

(ii) If $(k, H_{\mathsfsl{ext}})\in {\cal N}$, $k'\geq k$ and $H'_{ext}\geq H_{\mathsfsl{ext}}$ then $(k', \frac{k'}{k}H'_{ext})\in {\cal N}$.
\end{theorem}
The existence of such a Theorem is possible only because the system is invariant by homotheties (see, for example, \cite{merci-beaucoup-bernard} for the case of a superconductor restricted to a domain ${\cal D}$ of $\R^2$).

\begin{figure}
\begin{center}
\epsfig{figure=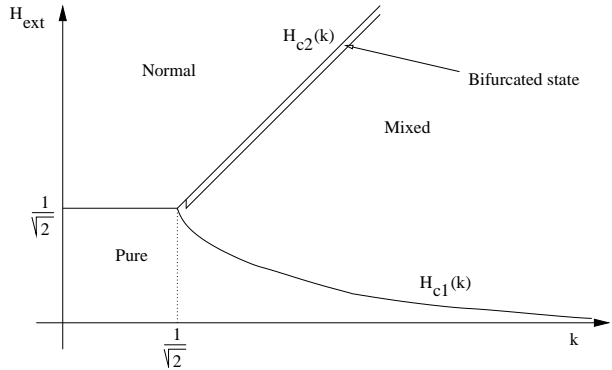, width=8cm}
\end{center}
\caption{Phase diagram in Abrikosov modelization}
\label{PhaseDiag}
\end{figure}

From this theorem we derived the existence of two functions $k\mapsto H_{c1}(k)$ and $k\mapsto H_{c2}(k)$ such that
\begin{equation*}
\begin{array}{rcl}
{\cal N}&=&\{(k, H_{\mathsfsl{ext}}), \mbox{~s.t.~} H_{\mathsfsl{ext}}\geq H_{c2}(k)\},\\
{\cal P}&=&\{(k, H_{\mathsfsl{ext}}), \mbox{~s.t.~} H_{\mathsfsl{ext}}\leq H_{c1}(k)\},\\
{\cal M}&=&\{(k, H_{\mathsfsl{ext}}), \mbox{~s.t.~} H_{c1}(k) < H_{\mathsfsl{ext}} < H_{c2}(k)\}\;.
\end{array}
\end{equation*}
Using this modelization we obtained in \cite{dutourII} the qualitative form of the phase diagram depicted in Figure \ref{PhaseDiag}, which is recalled in Section \ref{thePhaseDiagram}

This phase diagram is made of three curves:
\begin{enumerate}
\item[(i)] ({\em boundary normal-pure}) $H_{\mathsfsl{ext}}=H_{c1}(k)=H_{c2}(k)=\frac{1}{\sqrt{2}}$ with $k\leq \frac{1}{\sqrt{2}}$,
\item[(ii)] ({\em boundary normal-mixed}) $H_{\mathsfsl{ext}}=H_{c2}(k)=k$ with $k\geq \frac{1}{\sqrt{2}}$,
\item[(iii)] ({\em boundary pure-mixed}) $H_{\mathsfsl{ext}}=H_{c1}(k)$ with $k\geq \frac{1}{\sqrt{2}}$.
\end{enumerate}
The exact expression of curve (iii) is unknown.
Those three curves meet at the triple point $k=H_{\mathsfsl{ext}}=\frac{1}{\sqrt{2}}$. A key point of the proof is that the case $k=\frac{1}{\sqrt{2}}$ is exactly solvable thanks to the Bochner-Kodaira-Nakano formula explained in Section \ref{func}. Using a more advanced analysis of the case $k=\frac{1}{\sqrt{2}}$ in Section \ref{Quasi-Integrable-Case}, we prove in Section \ref{LOCAL-STUDY} the following Theorem:

\begin{theorem}\label{Main-theorem}
(i) There exist $\delta>0$ and $S>0$ such that for all $h$ in $[0, \delta]$, we have
\begin{equation*}
-h\leq H_{c1}(\frac{1}{\sqrt{2}}+h)-\frac{1}{\sqrt{2}}\leq -Sh\;.
\end{equation*}
(ii) The critical magnetic field $H_{c1}(k)$ is strictly decreasing at $k=\frac{1}{\sqrt{2}}$.
\end{theorem}

\section{THE CHANGE OF VARIABLE}\label{change-var}
In this Section, we recall the original formulation of the problem by V.Ginzburg and L.Landau in \cite{GL} and how it is related to our formulation.
They proposed the following expression for the density of energy in superconductors
\begin{equation*}
\frac{1}{2}\Vert i k^{-1}\mathbf{\nabla}\phi+\mathbf{A}\phi\Vert ^2+\frac{1}{4}(1-|\phi|^2)^2+\frac{1}{2}(\curl\,\mathbf{A}-H_{\mathsfsl{ext}})^2
\end{equation*}
This expression belongs to $L^1_{\mathsfsl{loc}}(\R^3)$ if $(\phi, \mathbf{A})$ is in the Sobolev space $H^1_{\mathsfsl{loc}}(\R^3; \C)\times H^1_{\mathsfsl{loc}}(\R^3; \R^3)$.
It is invariant under the gauge transform $(\phi', \mathbf{A}')=(\phi e^{ik g}, \mathbf{A}+\mathbf{\nabla} g)$ with $g\in H^2_{\mathsfsl{loc}}(\R^2)$;
this property is shared by other physically significant quantities like the density of superconducting electron $|\phi|^2$, the magnetic field $\curl\,\mathbf{A}$ and the current vector of superconducting electron $Re[\overline{\phi}(ik^{-1}\mathbf{\nabla}\phi+\mathbf{A}\phi)]$.

We assumed that the problem is invariant under translation along the z-axis. This means that we consider pairs $(\phi, \mathbf{A})$, which satisfies: for every $h\in\R$, the pair $(\phi, \mathbf{A})(x_1,x_2,x_3+h)$ is gauge equivalent to the pair $(\phi, \mathbf{A})$.

In fact, as proved in \cite{dutour} p.~17, we can assume that the pairs $(\phi, \mathbf{A})$ considered are independent of $x_3$ and satisfy $\mathbf{A}_{x_3}=0$. So, we can reduce ourself to a $2$-dimensional problem.

We take ${\cal L}$ a $2$-dimensional lattice of $\R^2$ with fundamental domain $\Omega$ of area $1$. We consider the dilated lattice : ${\cal L}_{\lambda}=\sqrt{\lambda}{\cal L}$ with fundamental domain $\Omega_{\lambda}=\sqrt{\lambda}\Omega$. 
Following Abrikosov, we choose $\lambda$ in $\R_+$ and restrict the analysis to pairs $(\phi, \mathbf{A})$, which are gauge periodic with respect to ${\cal L}_{\lambda}$ (\cite{Abrikosov}). This means that, for all $v\in{\cal L}_{\lambda}$, there exists $g^{v}\in H^2_{\mathsfsl{loc}}(\R^2)$ such that
\begin{equation*}
\phi(z+v)=e^{ikg^{v}(z)}\phi(z)\mbox{~~and~~}\mathbf{A}(z+v)=\mathbf{A}(z)+\mathbf{\nabla}g^{v}(z)\,\,\,.
\end{equation*}
Consequently, all the considered physical quantities are ${\cal L}_{\lambda}$-periodic. We denote by $|\Omega_{\lambda}|$ the area of $\Omega_{\lambda}$, which is actually equal to $\lambda$.

A classic consequence (see \cite{dutour}, \cite{Barany-Golubitsky}) of gauge periodicity is that there exist $d\in\Z$ satisfying to
\begin{equation*}\label{Quantization-equation}
2\pi d=k\int_{\Omega_{\lambda}}\curl\,\mathbf{A}\;.
\end{equation*}
We will then, according to Abrikosov, fix the quantization $d$ per unit cell equal to $1$.

The Ginzburg-Landau functional is obtained by integration of the local density over the fundamental domain $\Omega_{\lambda}$ and division by $|\Omega_{\lambda}|$. This gives:
\begin{equation*}\label{fonctionnelle-En2}
\begin{array}{rl}
F(\phi,\mathbf{A})&=\frac{1}{|\Omega_{\lambda}|}\int_{\Omega_{\lambda}}\frac{1}{2}\Vert i k^{-1}\mathbf{\nabla}\phi+\mathbf{A}\phi\Vert ^2+\frac{1}{4}(1-|\phi|^2)^2+\frac{1}{2}(\curl\,\mathbf{A}-H_{\mathsfsl{ext}})^2,
\end{array}
\end{equation*}
which should be understood as a mean energy.

We denote by $H_{\mathsfsl{int}}=\frac{1}{|\Omega_{\lambda}|}\int_{\Omega_{\lambda}}\curl\,\mathbf{A}$ the mean internal magnetic field induced by $\mathbf{A}$. 
The quantization relation is then rewritten as $2\pi=k\lambda H_{\mathsfsl{int}}$.

It is also a classical result (see \cite{Barany-Golubitsky}, \cite{yangyisong} or \cite{dutour}, p.~21-29) that we can associate to the pair $(\phi, \mathbf{A})$, another pair $(\phi', \mathbf{A}')$, with the same Ginzburg-Landau energy but satisfying to
\begin{itemize}
\item[(i)] $\mathbf{A}'=\frac{H_{\mathsfsl{int}}}{2\pi}\mathbf{A}_0+\mathbf{P}$
with $\mathbf{P}$ ${\cal L}_{\lambda}$-periodic, $\divergence\,\mathbf{P}=0$, $\int_{\Omega_{\lambda}}\mathbf{P}=0$,
\item[(ii)] $\phi'(z+v)=e^{ikg^{v}(z)}\phi'(z)$ with $g^{v}(x,y)=\frac{H_{\mathsfsl{int}}}{2}(v_{x}y-v_{y}x)$ for all $v\in {\cal L}_{\lambda}$.
\end{itemize}

This reduction is rather involved and is performed by a suitable gauge transform and a translation in $x,y$. The relation relating $\phi'(z+v)$ to $\phi'(z)$ actually defines the sections of a complex line bundle over the torus $\QuotS{\R^2}{\cal L}$; above result is so, a classification result.


With this expression one gets
\begin{equation*}\label{separation-des-termes}
\frac{1}{|\Omega_{\lambda}|}\int_{\Omega_{\lambda}}\frac{1}{2}(\curl\,\mathbf{A}-H_{\mathsfsl{ext}})^2=\frac{1}{|\Omega_{\lambda}|}\int_{\Omega_{\lambda}}\frac{1}{2}(\curl\,\mathbf{P})^2+\frac{1}{2}(H_{\mathsfsl{int}}-H_{\mathsfsl{ext}})^2 .
\end{equation*}
This leads to the simple expression $F(\phi,\mathbf{A})=F^{int}(\phi,\mathbf{P})+\frac{1}{2}(H_{\mathsfsl{int}}-H_{\mathsfsl{ext}})^2$ with
\begin{equation*}
\begin{array}{rcl}
F^{int}(\phi,\mathbf{P})&=&\frac{1}{\lambda}\int_{\Omega_{\lambda}}\frac{1}{2}\Vert i k^{-1}\mathbf{\nabla}\phi+\mathbf{A}\phi\Vert ^2+\frac{1}{4}(1-|\phi|^2)^2+\frac{1}{2}(\curl\,\mathbf{P})^2 .
\end{array}
\end{equation*}
The functional $F^{int}$ is called internal energy and depends only on $H_{\mathsfsl{int}}$, $k$, $\phi$ and $\mathbf{P}$. 

The quantities $H_{\mathsfsl{int}}$, $k$ and $\lambda$ are related by the quantization relation $2\pi=k\lambda H_{\mathsfsl{int}}$, which makes the analysis of $F^{int}$ cumbersome. 
So, we reduce the complexity of the computation by the following change of variables and of functions:
\begin{equation*}\label{change-of-variables}
\left\lbrace\begin{array}{rcl}
u(x)&=&\phi(x\sqrt{\frac{2\pi}{kH_{\mathsfsl{int}}}})\;,\\
\mathbf{a}(x)
&=&\sqrt{\frac{2\pi k}{H_{\mathsfsl{int}}}}[\mathbf{A}-\frac{H_{\mathsfsl{int}}}{2\pi}\mathbf{A}_0](x\sqrt{\frac{2\pi}{kH_{\mathsfsl{int}}}})=\sqrt{\frac{2\pi k}{H_{\mathsfsl{int}}}}\mathbf{A}(x\sqrt{\frac{2\pi}{kH_{\mathsfsl{int}}}})-\mathbf{A}_0(x)\;.
\end{array}\right.
\end{equation*}
We then obtain the formulation given in the introduction since the pair $(u,\mathbf{a})$ so defined belongs to ${\cal A}$ and verifies $E_{k,H_{\mathsfsl{int}}}(u,\mathbf{a})=F^{int}(\phi, \mathbf{P})$.

\section{THE FUNCTIONAL $E_{k,H_{\mathsfsl{int}}}$}\label{func}
\noindent Let us now analyze the functional $E_{k,H_{\mathsfsl{int}}}$ by assuming here that $k$ and $H_{\mathsfsl{int}}$ are fixed.

$E_{k,H_{\mathsfsl{int}}}$ is defined over ${\cal A}$ since $(u,\mathbf{a})$ of class $H^1$ guarantees local integrability of the density, while the compactness of the torus $\QuotS{\R^2}{\cal L}$ guarantees its integrability.

In fact, the variational theory of the functional $E_{k,H_{\mathsfsl{int}}}$ is easy (see \cite{dutour}) since the torus $\QuotS{\R^2}{\cal L}$ is compact and the non-linear partial differential equations obtained for the critical points are elliptic; the vector bundle adds only technical difficulties (see \cite{spin-geom}). More precisely one can prove successively that:
\begin{enumerate}
\item {\em Coerciveness}: for every $C\in\R$ there is a $C'>0$ such that $E_{k,H_{\mathsfsl{int}}}(u,\mathbf{a})<C$ implies $\Vert u\Vert_{H^1}+\Vert \mathbf{a}\Vert_{H^1}\leq C'$.
\item {\em Lower semicontinuity}: If $(u_n, \mathbf{a}_n) \in {\cal A}$ converges weakly to $(u, \mathbf{a})\in {\cal A}$, then $E_{k,H_{\mathsfsl{int}}}(u, \mathbf{a})\leq \underline{\lim}_n E_{k,H_{\mathsfsl{int}}}(u_n, \mathbf{a}_n)$.
\item {\em Minimum}: The functional $E_{k,H_{\mathsfsl{int}}}$ attains its minimum on at least one pair $(u,\mathbf{a})\in {\cal A}$.
\item {\em Ginzburg-Landau equations}: The minimizing pairs satisfy to the following equation
\begin{equation*}
\left\lbrace\begin{array}{rcl}
\mu[i\mathbf{\nabla}+\mathbf{A}_0+\mathbf{a}]^2u&=&(1-|u|^2)u\\
\Delta\mathbf{a}&=&\frac{1}{k^2}\Rez[\overline{u}(i\mathbf{\nabla}u+(\mathbf{A}_0+\mathbf{a})u)]
\end{array}\right.
\end{equation*}
\item {\em Regularity}: The pairs $(u,\mathbf{a})\in {\cal A}$ verifying the Ginzburg-Landau equations are in fact of class $C^{\infty}$.
\item {\em Maximum principle}: The pairs $(u, \mathbf{a})\in {\cal A}$ verifying the Ginzburg-Landau equations satisfy $|u|\leq 1$.
\end{enumerate}

We now explain the Bochner-Kodaira-Nakano formula for the functional $E_{k,H_{\mathsfsl{int}}}$ (see \cite{Demally}, \cite{jaffe-taubes} and \cite{wang-yang} for related formulas and results). 
This classical formula is also called Bogmol'nyi formula, Weitzenbock formula, Lichnerowicz formula (see \cite{Acker}) according to different scientific schools.

We set $\mathbf{C}=\mathbf{A}_0+\mathbf{a}$; we get $\curl\,\mathbf{C}=2\pi+\curl\,\mathbf{a}$ and define
\begin{equation*}\label{fonctionnelle-critique-2}
A_{+,H_{\mathsfsl{int}}}(u, \mathbf{a})=\int_{\Omega}\frac{\mu}{2}\vert D_{+}u\vert^2+\frac{1}{4}|\mu\curl\,\mathbf{C}-(1-|u|^2)|^2,
\end{equation*}
where $\mu=\frac{H_{\mathsfsl{int}}}{2\pi k}$ and $D_+=\frac{\partial }{\partial x}+i\frac{\partial }{\partial y}+C_{y} -iC_{x}$.

\begin{theorem}\label{Bochner-Kodaira-Nakano}
({\em Bochner-Kodaira-Nakano})\\
For all $(u, \mathbf{a})\in {\cal A}$, we have~:
\begin{equation*}\label{egalite-Bochner-Kodaira-Nakano}
E_{\frac{1}{\sqrt{2}},H_{\mathsfsl{int}}}(u, \mathbf{a})=\mu\pi-(\mu\pi)^2+A_{+,H_{\mathsfsl{int}}}(u, \mathbf{a}) .
\end{equation*}
\end{theorem}
\proof We perform computations with smooth functions and then extend by density. After expansion, simplification and regrouping one obtains 
\begin{equation*}\label{divergence-form}
\begin{array}{rcl}
\{A_{+,H_{\mathsfsl{int}}}-E_{\frac{1}{\sqrt{2}},H_{\mathsfsl{int}}}\}(u,\mathbf{a})
&=&\frac{1}{2}\int_{\Omega}\divergence\,\mathbf{W}-\mu\curl\,\mathbf{C}\\
&+&\frac{\mu^2}{4}\int_{\Omega}|\curl\,\mathbf{C}|^2-|\curl\,\mathbf{a}|^2
\end{array}
\end{equation*}
with $\mathbf{W}=\left(\begin{array}{c}
\overline{u}(i\frac{\partial u}{\partial y}+C_yu)\\
-\overline{u}(i\frac{\partial u}{\partial x}+C_xu)
\end{array}\right)$. The vector field $\mathbf{W}$ being ${\cal L}$-periodic, the integral of its divergence over $\Omega$ is $0$. The formula is then obtained by replacing $\curl\,\mathbf{C}$ by $2\pi+\curl\,\mathbf{a}$ and using $\int_{\Omega}\curl\,\mathbf{a}=0$. \qed


The {\em magnetic Schrodinger} operator is defined as $H=[i\mathbf{\nabla}+\mathbf{A}_{0}]^2$; its spectrum, called {\em Landau levels}, is recalled in next theorem.

\begin{theorem}
\begin{enumerate}
\item[(i)] The operator $H$ admits a self-adjoint extension over $L^2(E_1)$, also denoted by $H$, whose domain is $H^2(E_1)$.\\[-7mm]
\item[(ii)] It can be expressed as $H=L_+^*L_++2\pi$ with $[L_+,L_+^*]=4\pi$ and $L_+=2\partial_{\overline{z}}+\pi z$.

\item[(iii)] Its spectrum is discrete, $sp(H)=2\pi+4\pi\N$, and every eigenvalue is simple.\\[-7mm]
\item[(iv)] The eigenvector $u_0$ associated to $\lambda=2\pi$ satisfies $L_+(u_0)=0$ and has a unique simple zero in $\Omega$ denoted by $z_0$.
\end{enumerate}

\end{theorem}
\proof (i) and the discreteness of the spectrum follow from the fact that $H$ is an elliptic pseudo-differential operator of order $2$ defined over the vector bundle of a compact manifold (see \cite{spin-geom}).

Formula $H=L_+^*L_++2\pi$ and $[L_+, L_+^*]=4\pi$ are proved by first computing with smooth functions and then extending by density.

If we proved that the equation $L_+(u)=0$ has a unique solution $u_0$ up to scalar, then by the harmonic oscillator formalism we would get (iii).

In fact, if one writes, $u_0(z)=e^{-|z|^2\frac{\pi}{2}}s(z)$, then $s(z)$ is analytic. Furthermore, without loss of generality, we can assume that ${\cal L}$ is generated by the vectors $v_1=(u,0)$ and $v_2=(w,r)$ with $ru=1$. Then, after using gauge periodicity conditions, one finds the following expression for $u_0$:
\begin{equation*}
u_0(x,y)=e^{i\pi xy}\sum_{n\in\Z} e^{-\pi(y+nu)^2}e^{\pi n^2 iwu+2\pi nuix}\;.
\end{equation*}
This expression is a theta function; it is known that such function have a unique simple zero in $\Omega$ (see \cite{Chandrasekharan}). Another method of proof is the use of Rouch\'e Theorem as done in \cite{dutour}. \qed

\begin{theorem}\label{merveille-praguoise}
If $k\geq \frac{1}{\sqrt{2}}$ and $H_{\mathsfsl{int}}\geq k$, then $m_E(k,H_{\mathsfsl{int}})=\frac{1}{4}$. Furthermore, the minimum is met only by the pair $(0,0)$.
\end{theorem}
\proof We use following expansion of the functional $E_{k,H_{\mathsfsl{int}}}$:
\begin{equation*}\label{emerveillement-a-prague}
\begin{array}{rcl}
E_{k,H_{\mathsfsl{int}}}(u, \mathbf{a})
&\geq &E_{\frac{1}{\sqrt{2}},H_{\mathsfsl{int}}}(u, \mathbf{a})\\
&\geq &(\mu\pi)-(\mu\pi)^2+\int_{\Omega}\frac{\mu}{2}\vert D_{+}u\vert^2+\frac{1}{4}|2\mu\pi-1+\mu\curl\,\mathbf{a}+|u|^2|^2\\
&\geq &(\mu\pi)-(\mu\pi)^2+\frac{(2\mu\pi-1)^2}{4}+\frac{1}{4}\int_{\Omega}2(2\mu\pi-1)(\mu\curl\,\mathbf{a}+|u|^2)\\
&&+\frac{1}{4}\int_{\Omega}|\mu\curl\,\mathbf{a}+|u|^2|^2\\
&\geq &\frac{1}{4}+\frac{2\mu\pi-1}{2}\int_{\Omega}|u|^2\;.
\end{array}
\end{equation*}
Then using the hypothesis $2\mu\pi-1=\frac{H_{\mathsfsl{int}}}{k}-1\geq 0$, we get $m_E(k,H_{\mathsfsl{int}})\geq \frac{1}{4}$ by positivity of terms of above equation.

Now assume that $E_{k,H_{\mathsfsl{int}}}(u, \mathbf{a})=\frac{1}{4}$; in fact, last computation give us the following equalities:
\begin{equation*}
\left\lbrace\begin{array}{rclrcl}
0&=&(2\mu\pi-1)\int_{\Omega}|u|^2,    &0&=&\int_{\Omega}|\curl\,\mathbf{a}+|u|^2|^2 ,\\
0&=&(k^2-\frac{1}{2})\int_{\Omega}|\curl\,\mathbf{a}|^2,  &0&=&\int_{\Omega}\vert D_+u\vert^2 .
\end{array}\right.
\end{equation*}
The second equality give us $\curl\,\mathbf{a}+|u|^2=0$, which integrated over $\Omega$ yields
\begin{equation*}
\int_{\Omega}|u|^2=-\int_{\Omega}\curl\,\mathbf{a}=0 
\end{equation*}
and then $u=0$.

Now, using the equation $\divergence\,\mathbf{a}=0$, one obtains the equality $\curl^*\,\curl\,\mathbf{a}=\Delta\,\mathbf{a}=0$. The potential vector $\mathbf{a}$ is ${\cal L}$ periodic; so, it has to be constant. Now, the property $\int_{\Omega}\mathbf{a}=0$ yields $\mathbf{a}=0$. \qed

\section{THE PHASE DIAGRAM}\label{thePhaseDiagram}

Let us first consider the special case when $k=H_{\mathsfsl{ext}}=\frac{1}{\sqrt{2}}$. We have the following Lemma:
\begin{lemma}\label{lemma-preparatoire}
One has 

(i) $E_{\frac{1}{\sqrt{2}},\frac{1}{\sqrt{2}}}(H_{\mathsfsl{int}},u,\mathbf{a})=\frac{1}{4}+A_{+,H_{\mathsfsl{int}}}(u,\mathbf{a})$,

(ii) ${\cal E}_{\frac{1}{\sqrt{2}},\frac{1}{\sqrt{2}}}=\frac{1}{4}$,

(iii) $(\frac{1}{\sqrt{2}},\frac{1}{\sqrt{2}})\in {\cal P}\cap {\cal N}$.
\end{lemma}
\proof (i) is in fact a rewriting of the Bochner-Kodaira-Nakano formula; it yields (ii) by positivity of $A_{+, H_{\mathsfsl{int}}}$, while (iii) is obtained by remarking that $E_{\cal N}=\frac{1}{4}=\frac{1}{2}(\frac{1}{\sqrt{2}})^2=E_{\cal P}$. \qed


\begin{theorem}\label{usage-theoreme-monotonie}
{\it (Type I superconductors)} If $k\leq \frac{1}{\sqrt{2}}$, then:
\begin{enumerate}
\item[(i)] If $H_{\mathsfsl{ext}}\leq \frac{1}{\sqrt{2}}$, then ${\cal E}_{k,H_{\mathsfsl{ext}}}=E_{\cal P}$ and $(k, H_{\mathsfsl{ext}})\in {\cal P}$,
\item[(ii)] If $H_{\mathsfsl{ext}}\geq \frac{1}{\sqrt{2}}$, then ${\cal E}_{k,H_{\mathsfsl{ext}}}=E_{\cal N}$ and $(k, H_{\mathsfsl{ext}})\in {\cal N}$.
\end{enumerate}

\end{theorem}
\proof Lemma \ref{lemma-preparatoire} combined with Theorem \ref{TheoremMonotony}.(i) give the result in the case $H_{\mathsfsl{ext}}\leq \frac{1}{\sqrt{2}}$.

In particular, if $k\leq \frac{1}{\sqrt{2}}$ we have $(k, \frac{1}{\sqrt{2}})\in{\cal P}$ and so, ${\cal E}_{k, \frac{1}{\sqrt{2}}}=E_{\cal P}=\frac{1}{2}(\frac{1}{\sqrt{2}})^2=\frac{1}{4}=E_{\cal N}$; therefore Theorem \ref{TheoremMonotony}.(ii) gives the conclusion in case $H_{\mathsfsl{ext}}\geq \frac{1}{\sqrt{2}}$. \qed

\begin{theorem}\label{usage-theoreme-monotonie-2}
{\it (Type II superconductors)} If $H_{\mathsfsl{ext}}\geq k\geq \frac{1}{\sqrt{2}}$, then:
\begin{enumerate}
\item[(i)] If $H_{\mathsfsl{ext}}\geq k$, then ${\cal E}_{k,H_{\mathsfsl{ext}}}=E_{\cal N}$ and $(k, H_{\mathsfsl{ext}})\in {\cal N}$,
\item[(ii)] If $H_{\mathsfsl{ext}}< k$, then $(k, H_{\mathsfsl{ext}})\notin {\cal N}$.
\end{enumerate}

\end{theorem}
\proof Lemma \ref{lemma-preparatoire} combined with Theorem~\ref{TheoremMonotony}.(ii) gives (i).

By setting $H_{\mathsfsl{int}}=H_{\mathsfsl{ext}}$, $u=\alpha u_0$, $\mathbf{a}=0$
and doing a development of order $2$ around the pair $(0,0)$, one obtains
\begin{equation*}
E_{k,H_{\mathsfsl{ext}}}(H_{\mathsfsl{ext}},\alpha u_0,0)=E_{k,H_{\mathsfsl{ext}}}(\alpha u_0,0)=\frac{1}{4}+\frac{1}{2}(\frac{H_{\mathsfsl{ext}}}{k}-1)\alpha^2+o(\alpha^2)\;.
\end{equation*}
Since $k > H_{\mathsfsl{ext}}=H_{\mathsfsl{int}}$, one obtains for $\alpha$ small $E_{k,H_{\mathsfsl{ext}}}(H_{\mathsfsl{ext}},\alpha u_0,0)<\frac{1}{4}$; so, the energy will be lower than $\frac{1}{4}$, i.e. $(k,H_{\mathsfsl{ext}})\notin {\cal N}$. \qed

\section{ANALYSIS OF THE CASE $k=\frac{1}{\sqrt{2}}$}\label{Quasi-Integrable-Case}
\noindent In this section we will find all pairs $(u, \mathbf{a})$ verifying $A_{+,H_{\mathsfsl{int}}}(u, \mathbf{a})$, thus get the value of $m_E(\frac{1}{\sqrt{2}}, H_{\mathsfsl{int}})$. A similar study is done in \cite{almogII} for a rectangular problem. 
In book \cite{jaffe-taubes}, the case considered is of $u$ defined over $\R^2$, while in paper (\cite{Garcia-prada}) the problem is considered over a Riemann surface. Also, in \cite{jaffe-taubes} it is proved that all critical points of the Ginzburg-Landau functional are solution of the Bogmol'nyi equations, but their proof does not apply to our case.

The papers (\cite{Kazdan-Warner}), (\cite{caffarelli}), (\cite{wang-yang}) are devoted to existence theorem concerning the Kazdan-Warner equation. They get as a byproduct existence Theorems for the self-dual equations.

\begin{theorem}\label{Kazdan-Warner-Salamon-Auroux}
({\it Kazdan-Warner}, see \cite{Kazdan-Warner}) If $h$ is a positive function, $h\not= 0$, and $C^{\infty}(\Tore{\cal L})$. If $A>0$ then the equation
\begin{equation*}\label{equation-Kazdan-Warner}
-\Delta f+e^f h=A
\end{equation*}
has a unique solution $f$ in $C^{\infty}(\Tore{\cal L})$.
\end{theorem}

We define
\begin{equation*}\label{definition-solution-dirac}
\left\lbrace\begin{array}{rcl}
u_{H_{\mathsfsl{int}}}&=&u_0e^{f_{H_{\mathsfsl{int}}}}\\
\mathbf{a_{H_{\mathsfsl{int}}}}&=&(\frac{\partial f_{H_{\mathsfsl{int}}}}{\partial y},-\frac{\partial f_{H_{\mathsfsl{int}}}}{\partial x}),
\end{array}\right.
\end{equation*}
with $f_{H_{\mathsfsl{int}}}$ being the unique solution of $1-2\mu\pi=|u_0|^2e^{2f}-\mu\Delta\,f$ and $\mu=\frac{H_{\mathsfsl{int}}}{\pi \sqrt{2}}$.

Let us introduce first the following family of sections of $E_1$:
\begin{equation*}
u_h(x,y)=e^{i\pi(h_y x-h_x y)}u_0(z-h)\;.
\end{equation*}
Recall that $z_0$ is the zero of $u_0$ in $\QuotS{\R^2}{\cal L}$; the section $u_h$ verifies the following easy properties 
\begin{equation*}\label{propriete-phi_h}
\left\lbrace\begin{array}{cc}
u_h\in C^{\infty}(E_1),        &L_+(u_h)=2\pi hu_h,\\
\multicolumn{2}{c}{u_h(z)=0\mbox{~if~and~only~if~}z\in z_0+h+{\cal L}\;\;.}
\end{array}\right.
\end{equation*}
Furthermore, for any $h\in\R^2$, $v\in {\cal L}$, there exists $\alpha\in\R$ such that
\begin{equation*}
u_{h+v}(z)=e^{i\alpha}e^{2i\pi(v_y x-v_x y)}u_h(z)\;.
\end{equation*}

\begin{theorem}\label{Solution_Set}
We assume $H_{\mathsfsl{int}}\leq \frac{1}{\sqrt{2}}$. 

(i) If $(u,\mathbf{a})\in {\cal A}$ satisfies $A_{+,H_{\mathsfsl{int}}}(u,\mathbf{a})=0$, then there exist $c\in\R$ such that $(u,\mathbf{a})=(e^{ic}u_{H_{\mathsfsl{int}}}, \mathbf{a_{H_{\mathsfsl{int}}}})$.

(ii) The pair $(u_{H_{\mathsfsl{int}}}, \mathbf{a_{H_{\mathsfsl{int}}}})$ satisfies to
\begin{equation*}\label{FundamentalFormulas}
\left\lbrace\begin{array}{l}
\int_{\Omega}(1-|u_{H_{\mathsfsl{int}}}|^2)^2=\mu^2[(2\pi)^2+\int_{\Omega}|\curl\,\mathbf{a_{H_{\mathsfsl{int}}}}|^2]\\
\int_{\Omega}\frac{\mu}{2}\Vert i\mathbf{\nabla}u_{H_{\mathsfsl{int}}}+(\mathbf{A}_0+\mathbf{a})u_{H_{\mathsfsl{int}}}\Vert^2+\frac{\mu^2}{2}|\curl\,\mathbf{a}_{H_{\mathsfsl{int}}}|^2=(\mu\pi)-2(\mu\pi)^2\;.
\end{array}\right.
\end{equation*}

\end{theorem}
\proof Let $(u, \mathbf{a})\in {\cal A}$ be a pair satisfying $A_{+,H_{\mathsfsl{int}}}(u, \mathbf{a})=0$, it then verifies the following Bogmol'nyi equations
\begin{equation*}\label{equation-bogolomyi}
D_{+}u=L_+u+(a_y-ia_x)u=0\mbox{~~and~~}2\mu\pi+\mu\curl\,\mathbf{a}=1-|u|^2
\end{equation*}
and, by Theorem~\ref{Bochner-Kodaira-Nakano}, minimizes the functional $E_{\frac{1}{\sqrt{2}},H_{\mathsfsl{int}}}$.
Therefore, by Section \ref{func} it satisfies the Ginzburg-Landau equations and so, it is $C^{\infty}$.

Since the vector bundle $E_1$ is non trivial the section $u$ possess at least one zero in $\QuotS{\R^2}{\cal L}$, which we write as $z_h=z_0+h$.

The zero-set of the function $u$ defined on $\R^2$ contains $z_h+{\cal L}$, while the zero-set of $u_h$ is exactly $z_h+{\cal L}$; so, one defines on $\R^2-(z_h+{\cal L})$ the function
\begin{equation*}
f=\frac{u}{u_h}\; .
\end{equation*}
Since both $u$ and $u_h$ are section of the vector bundle $E_1$, the function $f$ is ${\cal L}$-periodic. The equation $D_+u=0$ is rewritten on $\R^2-(z_h+{\cal L})$ as:
\begin{equation*}
0=2(\partial_{\overline{z}}f)u_h+fD_+u_h=2(\partial_{\overline{z}}f)u_h+[2\pi hf+(a_y-ia_x)f]u_h\;,
\end{equation*}
Since $u_h$ is not zero on $\R^2-(z_h+{\cal L})$ we obtain:
\begin{equation*}\label{equation-z-barre}
\partial_{\overline{z}}f=fw\mbox{~~with~~}w=\frac{1}{2}[(-a_y-2\pi h_x)+i(a_x-2\pi h_y)]\;.
\end{equation*}
Note that the function $w$ is defined on $\R^2$, also it is $C^{\infty}$ and ${\cal L}$-periodic.

We now want to extend $f$ to $\R^2$: it is a classic result of complex analysis that the equation $\partial_{\overline{z}}k=w$ has a $C^{\infty}$ solution $k$ on $\R^2$.

The function $g=fe^{-k}$ is defined on $\R^2 -(z_h+{\cal L})$, satisfies $\partial_{\overline{z}}g=0$ and is so, analytic.
If $m\in z_h+{\cal L}$ then $u=O(z-m)$, since $u$ is $C^{\infty}$. The complex $m$ is a simple zero of $u_h$, consequently $u_h^{-1}=O(|z-m|^{-1})$ and $f=O(1)$ at $m$.

The function $g$ stay bounded around $m$ and is analytic outside $m$. By a classic result of complex analysis, we get that $g$ can be extended to $m$ in a complex analytic function. The function $g$ is extended to $\C$ and so, $f$ too.

The function $g$ is analytic and so, its zero set is discrete.
There exist a translate $\Omega'$ of $\Omega$
such that the boundary $\partial\Omega'$ of $\Omega'$ does not meet any zero of $g$.

By Rouché theorem the number $n$ of zero of $g$ in $\Omega'$ is equal to :
\begin{equation*}
n=\frac{1}{2\pi i}\int_{\partial\Omega'}\frac{\partial_z g}{g}dz
=\frac{1}{2\pi i}\int_{\partial\Omega'}\frac{\partial_z f}{f}dz-\frac{1}{2\pi i}\int_{\partial\Omega'}\partial_z kdz=\frac{-1}{2\pi i}\int_{\partial\Omega'}\partial_z kdz .
\end{equation*}
The integral of $\frac{\partial_{z}f}{f}$ over $\partial\Omega'$ is zero, since $f$ is ${\cal L}$-periodic.

Now using Stokes theorem, we get :
\begin{equation*}
\begin{array}{rcl}
n&=&\frac{-1}{2\pi i}\int_{\partial \Omega'}\partial_z kdz
=\frac{-1}{2\pi i}\int_{\Omega'}d(\partial_z kdz)
=\frac{-1}{2\pi i}\int_{\Omega'}\partial_{\overline{z}}\partial_z k d\overline{z}\wedge dz\\[5mm]
&=&\frac{-1}{2\pi i}\int_{\Omega'}\partial_z\partial_{\overline{z}}k d\overline{z}\wedge dz
=\frac{-1}{2\pi i}\int_{\Omega'}\partial_z w d\overline{z}\wedge dz .
\end{array}
\end{equation*}
The function $w$ is ${\cal L}$-periodic; consequently the function $\partial_z w$ is a ${\cal L}$-periodic function, which has integral zero over $\Omega'$; so, $n=0$.

Since $f=ge^{k}$, the function $f$ has no zero over $\R^2$. Since $\R^2$ is simply connected there exist a complex valued $C^{\infty}$ function $\psi$ such that $f=e^{\psi}$.

The function $\psi$ is not ${\cal L}$-periodic, but since the function $f$ is ${\cal L}$-periodic and $C^{\infty}$ there exist two integer $n_1, n_2$ such that
\begin{equation*}
\psi(z+v_1)=\psi(z)+2\pi i n_1\mbox{~~and~~}\psi(z+v_2)=\psi(z)+2\pi i n_2\; .
\end{equation*}
We pose $v'=n_1 v_2-n_2v_1$, the function $\psi_2(z)=\psi(z)-2\pi i\det(z,v')$ is ${\cal L}$-periodic, and we have :
\begin{equation*}
u(z)=f(z)u_h(z)=e^{\psi_2(z)+2\pi i[x v'_y-y v'_x]}u_h(z)=e^{\psi_2(z)-i\alpha}u_{h+v'}(z)
\end{equation*}
with $\alpha\in\R$. We set $h_3=h+v'$, $\psi_3=\psi_2-i\alpha$, and we rewrite $u$ as:
\begin{equation*}
u(z)=e^{\psi_3(z)}u_{h_3}(z)
\end{equation*}
with $\psi_3$ a ${\cal L}$-periodic $C^{\infty}$ function.
The Bogmol'nyi equations are rewritten as
\begin{equation*}
\left\lbrace\begin{array}{rcl}
\frac{\partial \psi_3}{\partial \overline{z}}&=&\frac{1}{2}[(-a_y-\pi h_{3,x})+i(a_x-\pi h_{3,y})]\;,\\
0&=&2\mu\pi-1+|u_{h_3}|^2e^{2\Rez\,\psi_3}+\mu\curl\,\mathbf{a} .
\end{array}\right.
\end{equation*}
The real and imaginary part of first equation give us the expression of the potential vector:
\begin{equation*}
\left\lbrace\begin{array}{rcl}
a_x&=&\pi h_{3,y}+\frac{\partial \Rez\,\psi_3}{\partial y}+\frac{\partial \Imz\,\psi_3}{\partial x}\;,\\
a_y&=&-\pi h_{3,x}-\frac{\partial \Rez\,\psi_3}{\partial x}+\frac{\partial \Imz\,\psi_3}{\partial y}\;.
\end{array}\right.
\end{equation*}
The equation $\divergence\,\mathbf{a}=0$ is then rewritten as $\Delta\,\Imz\,\psi_3=0$. Thus $\Imz\,\psi_3$ is constant, since it is ${\cal L}$-periodic.
We now write $\psi_3=f+ic$ with $f$ a real $C^{\infty}$, ${\cal L}$-periodic function; so, one has
\begin{equation*}
a_x=\pi h_{3,y}+\frac{\partial f}{\partial y}\mbox{~~and~~}a_y=-\pi h_{3,x}-\frac{\partial f}{\partial x}\; .
\end{equation*}
The functions $\mathbf{a}$, $\frac{\partial f}{\partial x}$, and $\frac{\partial f}{\partial y}$ have zero integral over $\Omega$. So, we have $h_3=0$ and the zero of $u$ in $\Omega$ is $z_0$.

One then obtain $\curl\,\mathbf{a}=-\Delta\,f$ and the following equation for $f$:
\begin{equation*}
0=2\mu\pi-1+|u_0|^2e^{2f}-\mu\Delta\,f\; .
\end{equation*}
So, one gets $f=f_{H_{\mathsfsl{int}}}$; now above equation rewrites as
\begin{equation*}
-\mu\curl\,\mathbf{a_{H_{\mathsfsl{int}}}}=2\mu\pi-1+|u_{H_{\mathsfsl{int}}}|^2\; .
\end{equation*}
It yields $\int_{\Omega}|u_{H_{\mathsfsl{int}}}|^2=1-2\mu\pi$ and 
$\int_{\Omega}(1-|u_{H_{\mathsfsl{int}}}|^2)^2=\mu^2[(2\pi)^2+\int_{\Omega}|\curl\,\mathbf{a_{H_{\mathsfsl{int}}}}|^2]$, the second equation of (ii) is then obtained by Theorem \ref{Bochner-Kodaira-Nakano}. \qed

\begin{corollary}.\label{calculs-energie}
For every positive $H_{\mathsfsl{int}}$ one has:
\begin{equation*}
m_E(\frac{1}{\sqrt{2}}, H_{\mathsfsl{int}})=\left\lbrace\begin{array}{lcl}
\frac{H_{\mathsfsl{int}}}{\sqrt{2}}-(\frac{H_{\mathsfsl{int}}}{\sqrt{2}})^2&\mbox{~if~}&H_{\mathsfsl{int}}\leq \frac{1}{\sqrt{2}}\;,\\
\frac{1}{4}&\mbox{~if~}&H_{\mathsfsl{int}}\geq \frac{1}{\sqrt{2}}\;.
\end{array}\right.
\end{equation*}
\end{corollary}
\proof By Theorem~\ref{Bochner-Kodaira-Nakano}, one has the inequality $m_E(\frac{1}{\sqrt{2}},H_{\mathsfsl{int}})\geq \frac{H_{\mathsfsl{int}}}{\sqrt{2}}-(\frac{H_{\mathsfsl{int}}}{\sqrt{2}})^2$, since $A_{+,H_{\mathsfsl{int}}}\geq 0$. This lower bound is attained by the pair $(u_{H_{\mathsfsl{int}}}, \mathbf{a}_{H_{\mathsfsl{int}}})$.

Theorem~\ref{merveille-praguoise} give the result if $H_{\mathsfsl{int}}\geq \frac{1}{\sqrt{2}}$. \qed

\begin{remark}
It can be shown that the pair $(u_{H_{\mathsfsl{int}}}, \mathbf{a_{H_{\mathsfsl{int}}}})$ depends continuously on $H_{\mathsfsl{int}}$ and vanish for $H_{\mathsfsl{int}}=\frac{1}{\sqrt{2}}$, i.e. it is a {\em bifurcated state} (see \cite{dutour}).
\end{remark}

\section{LOCAL STUDY}\label{LOCAL-STUDY}
We define 
\begin{equation*}
\begin{array}{rcl}
H_k(u,\mathbf{a})&=&\frac{1}{4\pi k}\int_{\Omega}\Vert i\mathbf{\nabla}u+(\mathbf{A}_0+\mathbf{a})u\Vert^2\\
&+&\sqrt{[\frac{1}{2}+\frac{1}{2(2\pi)^2}\int_{\Omega}|\curl\,\mathbf{a}|^2][\int_{\Omega}(1-|u|^2)^2]}\;.
\end{array}
\end{equation*}

\begin{theorem}\label{ExpressionCriticalField}
If $k\geq \frac{1}{\sqrt{2}}$ then $H_{c1}(k)=\inf_{(u,\mathbf{a})\in{\cal A}}H_k(u,\mathbf{a})$.
If this infimum is attained on a pair, say, $(u', \mathbf{a}')\in {\cal A}$, then one has
\begin{equation*}
E_{k,H_{c1}(k)}(H_{\mathsfsl{int}}, u',\mathbf{a}')=\frac{H^2_{c1}(k)}{2}
\mbox{~~with~~}
H_{\mathsfsl{int}}=\frac{1}{2}\sqrt{\frac{\int_{\Omega}(1-|u'|^2)^2}{\frac{1}{2}+\frac{1}{2(2\pi)^2}\int_{\Omega}|\curl\,\mathbf{a'}|^2}}\;.
\end{equation*}

\end{theorem}
\proof By Section \ref{Introduction}, we have $(k, H_{\mathsfsl{ext}})\in {\cal P}$ equivalent to:
\begin{equation*}
E_{k,H_{\mathsfsl{int}}}(u,\mathbf{a})+\frac{1}{2}(H_{int}-H_{ext})^2\geq \frac{H_{ext}^2}{2}\;,
\end{equation*}
which after simplification is equivalent to
\begin{equation*}
\left\lbrace\begin{array}{l}
H_{\mathsfsl{int}}[\frac{1}{2}+\frac{1}{2(2\pi)^2}\int_{\Omega}|\curl\,\mathbf{a}|^2]+\frac{1}{4H_{\mathsfsl{int}}}\int_{\Omega}(1-|u|^2)^2\\
+\frac{1}{4\pi k}\int_{\Omega}\Vert i\mathbf{\nabla}u+(\mathbf{A}_0+\mathbf{a})u\Vert^2\geq H_{ext} .
\end{array}\right. 
\end{equation*}
The minimum over $H_{\mathsfsl{int}}>0$ of the above expression is attained for $H_{\mathsfsl{int}}=\frac{1}{2}\sqrt{\frac{\int_{\Omega}(1-|u|^2)^2}{\frac{1}{2}+\frac{1}{2(2\pi)^2}\int_{\Omega}|\curl\,\mathbf{a}|^2}}$ which yields the Theorem. \qed

The above expression of $H_{c1}(k)$ allow us to obtain $H_{c1}(k)=O(\frac{\ln\,k}{k})$ (see \cite{dutour}).
From Theorem~\ref{usage-theoreme-monotonie}, one has $H_{c1}(\frac{1}{\sqrt{2}})=\frac{1}{\sqrt{2}}$.
\begin{theorem}
The set of pairs $(u, \mathbf{a})\in {\cal A}$ verifying $H_{\frac{1}{\sqrt{2}}}(u, \mathbf{a})=\frac{1}{\sqrt{2}}$ is
\begin{equation*}
(e^{ic}u_{H_{\mathsfsl{int}}}, \mathbf{a_{H_{\mathsfsl{int}}}})
\end{equation*}
with $c\in\R$ and $0<H_{\mathsfsl{int}}\leq \frac{1}{\sqrt{2}}$.
\end{theorem}
\proof If $(u,\mathbf{a})\in {\cal A}$ satisfies $H_{\frac{1}{\sqrt{2}}}(u,\mathbf{a})=\frac{1}{\sqrt{2}}$, then one has
\begin{equation*}
E_{\frac{1}{\sqrt{2}},\frac{1}{\sqrt{2}}}(H_{\mathsfsl{int}}, u, \mathbf{a})=\frac{1}{4}
\mbox{~~and~~}
H_{\mathsfsl{int}}=\frac{1}{2}\sqrt{\frac{\int_{\Omega}(1-|u|^2)^2}{\frac{1}{2}+\frac{1}{2(2\pi)^2}\int_{\Omega}|\curl\,\mathbf{a}|^2}}\;.
\end{equation*}
By Lemma \ref{lemma-preparatoire}.(i), first equation simplifies to $A_{+,H_{\mathsfsl{int}}}(u,\mathbf{a})=0$, and then using Theorem~\ref{Solution_Set} to $(u,\mathbf{a})=(e^{ic}u_{H_{\mathsfsl{int}}}, \mathbf{a_{H_{\mathsfsl{int}}}})$.

When the expression of $(u,\mathbf{a})$ is substituted into the second equation, one obtains
\begin{equation*}
4H^2_{\mathsfsl{int}}
=\frac{\int_{\Omega}(1-|u_{H_{\mathsfsl{int}}}|^2)^2}{\frac{1}{2}+\frac{1}{2(2\pi)^2}\int_{\Omega}|\curl\,\mathbf{a_{H_{\mathsfsl{int}}}}|^2}\;.
\end{equation*}
By Theorem~\ref{Solution_Set}.(ii), this relation is always satisfied. \qed

\begin{theorem}
(i) There exist $\delta>0$ and $S>0$ such that for all $h$ in $[0, \delta]$, we have
\begin{equation*}
-h\leq H_{c1}(\frac{1}{\sqrt{2}}+h)-\frac{1}{\sqrt{2}}\leq -Sh\;.
\end{equation*}
(ii) The critical magnetic field $H_{c1}(k)$ is strictly decreasing at $k=\frac{1}{\sqrt{2}}$.
\end{theorem}
\proof The expression of $H_{c1}(k)$ obtained in Theorem~\ref{ExpressionCriticalField} give us that the function $k\mapsto kH_{c_1}(k)$ is increasing; this yields the lower bound.

Now we will prove the upper bound by using the $(u_{H_{\mathsfsl{int}}}, \mathbf{a_{H_{\mathsfsl{int}}}})$ as quasimodes.
If $k=\frac{1}{\sqrt{2}}+h$ then we will have
\begin{equation*}
H_k(u_{H_{\mathsfsl{int}}}, \mathbf{a_{H_{\mathsfsl{int}}}})=\frac{1}{\sqrt{2}}
-\frac{h}{2\pi}\int_{\Omega}\Vert i\mathbf{\nabla}u_{H_{\mathsfsl{int}}}+(\mathbf{A}_0+\mathbf{a_{H_{\mathsfsl{int}}}})u_{H_{\mathsfsl{int}}}\Vert^2+o(h)\;.
\end{equation*}
We get the following values of $S$ using Bochner-Kodaira-Nakano
\begin{equation*}
\begin{array}{rcl}
S&=&\sup_{0<H_{\mathsfsl{int}}<\frac{1}{\sqrt{2}}}\frac{1}{2\pi}\int_{\Omega}\Vert i\mathbf{\nabla}u_{H_{\mathsfsl{int}}}+(\mathbf{A}_0+\mathbf{a_{H_{\mathsfsl{int}}}})u_{H_{\mathsfsl{int}}}\Vert^2\\
 &=&\sup_{0<H_{\mathsfsl{int}}<\frac{1}{\sqrt{2}}}[1-\frac{H_{\mathsfsl{int}}}{\frac{1}{\sqrt{2}}}-\frac{H_{\mathsfsl{int}}}{2\pi^2\sqrt{2}}\int_{\Omega}|\curl\, \mathbf{a_{H_{\mathsfsl{int}}}}|^2]
\end{array}
\end{equation*}
follows from Theorem~\ref{Solution_Set}.(ii). \qed

One may want now to know the exact value of $S$ at $\frac{1}{\sqrt{2}}$.  Using numerical simulations we obtain that the function 
\begin{equation*}
\chi(H_{\mathsfsl{int}})=1-\sqrt{2}H_{\mathsfsl{int}}-\frac{H_{\mathsfsl{int}}}{2\pi^2\sqrt{2}}\int_{\Omega}|\curl\, \mathbf{a_{H_{\mathsfsl{int}}}}|^2
\end{equation*}
is decreasing and has a limit of approximately $0.78$ at $H_{\mathsfsl{int}}=0$ for a square lattice.

\end{document}